\documentclass[onecolumn,nofootinbib]{revtex4}
\usepackage{graphicx}
\usepackage{latexsym}
\usepackage{tabularx}

\begin{document}

\hfill USTC-ICTS-13-11

\title{Constraints on Cosmological Models from Hubble Parameters Measurements}

\author{Wei Zheng${}^{a,b}$}
\author{Hong Li${}^{b,c}$}
\author{Jun-Qing Xia${}^{b}$}
\author{You-Ping Wan${}^{d}$}
\author{Si-Yu Li${}^{d}$}
\author{Mingzhe Li${}^{e}$}

\affiliation{${}^a$Department of Physics, Nanjing University, Nanjing 210093, China}
\affiliation{${}^b$Key Laboratory of Particle Astrophysics, Institute of High Energy Physics, Chinese Academy of Science, P. O. Box 918-3, Beijing 100049, P. R. China}
\affiliation{${}^c$National Astronomical Observatories, Chinese Academy of Sciences, Beijing 100012, P. R. China}
\affiliation{${}^d$Theoretical Physics Division, Institute of High Energy Physics, Chinese Academy of Science, P. O. Box 918-4, Beijing 100049, P. R. China}
\affiliation{${}^e$Interdisciplinary Center for Theoretical Study, University of Science and Technology of China, Hefei, Anhui 230026, China}

\date{\today}

\begin{abstract}

In this paper, we study the cosmological constraints from the measurements of Hubble parameters---$H(z)$ data. Here, we consider two kinds of $H(z)$ data: the direct $H_0$ probe from the Hubble Space Telescope (HST) observations of Cepheid variables with $H_0=73.8\pm2.4$ ${\rm km\,s^{-1}\,Mpc^{-1}}$ and several measurements on the Hubble parameter at high redshifts $H(z)$. Employing Markov Chain Monte Carlo method, we also combine the WMAP nine-year data (WMAP9), the baryon acoustic oscillations (BAO) and type Ia supernovae (SNIa) ¡®¡®Union2.1¡¯¡¯ compilation to determine the cosmological parameters, such as the equation of state (EoS) of dark energy $w$, the curvature of the universe $\Omega_k$, the total neutrino mass $\sum{m_\nu}$, the effective number of neutrinos $N_{\rm eff}$, and the parameters associated with the power spectrum of primordial fluctuations. These $H(z)$ data provide extra information on the accelerate rate of our Universe at high redshifts. Therefore, adding these $H(z)$ data significantly improves the constraints on cosmological parameters, such as the number of relativistic species. Moreover, we find that direct prior on $H_0$ from HST can also give good constraints on some parameters, due to the degeneracies between these parameters and $H_0$.

\end{abstract}

\maketitle

\section{Introduction}\label{Int}

Cosmological measurements, such as the cosmic microwave background radiation (CMB), type Ia supernovae (SNIa), large scale structure (LSS), as well as baryon acoustic oscillation (BAO), play a crucial role in our understanding of the universe and also in constraining the cosmological parameters. However, due to various degeneracies among these cosmological parameters \cite{degeneracy}, even the high resolution, full sky, maps of the CMB temperature anisotropies from WMAP \cite{wmap} and Planck \cite{planck_map} and hundreds of samples of SNIa data at low redshifts \cite{union2} can not give very good constraints on some crucial parameters, such as the equation of state (EoS) of dark energy and the total neutrino mass. Therefore, in order to break degeneracies among these parameters, we have to consider some additional observational information to improve the constraints (see e.g. refs. \cite{grb,weaklensing,isw_li}). In this regard, exploring new probes has great importance.

The measurements of Hubble parameters can potentially to be a complementary probe in constraining cosmological parameters. The Hubble parameter, defined as $H(z)=\dot{a}/{a}$ where $a$ is the scale factor of the universe, characterizes the expansion rate of our universe at different redshifts, and depends on the differential age of the universe as a function of redshift
\begin{equation}
H(z)=-\frac{1}{1+z}\frac{dz}{dt}~.
\end{equation}
Therefore, measuring the $dz/dt$ could straightforwardly estimate $H(z)$, which was firstly proposed by ref. \cite{jimenez}. They selected samples of passively evolving galaxies with high-quality spectroscopy, and then used stellar population models to constrain the age of the oldest stars in these galaxies. After that, they computed differential ages at different redshifts and obtained the determinations of Hubble parameter \cite{simon,stern,zhang2012,moresco}. Moreover, the Hubble parameter can also be obtained from the BAO measurement. By observing the typical acoustic scale in the light-of-sight direction, one can extract the expansion rate of the universe at certain redshift. Ref. \cite{gaztanaga} analyzed the information of Hubble parameter at redshift $z=0.24$ and $z=0.43$ from the SDSS DR6 and DR7 data. Recently, these $H(z)$ data have been widely used on the determination of cosmological parameters, such as the effective number of neutrinos \cite{verde,riemer}, the EoS of dark energy \cite{lazkoz,zhu,ratra}, the cosmography scenario \cite{cosmo_v,cosmo_xia}, the modified gravity models \cite{bengochea,aviles,tjzhang}.

Besides these $H(z)$ data, we also have the direct probe on the current Hubble constant $H_0$. Using the observations of nearby SNIa samples, the Hubble Space Telescope (HST) collaboration provides the direct measurement on $H_0$ with high precision, namely the $68\%$ C.L. limit $H_0=73.8\pm2.4$ ${\rm km\,s^{-1}\,Mpc^{-1}}$ \cite{hst_riess} or $H_0=74.3\pm1.5{\rm (stat.)}\pm2.1{\rm (sys.)}$ ${\rm km\,s^{-1}\,Mpc^{-1}}$ \cite{hst_freedman}.

In this paper, we summarize the recent $H(z)$ data and investigate their constraining power on the determination of cosmological parameters. Combining with the WMAP nine-year data (WMAP9), the ``Union 2.1'' compilation SNIa sample and the several recent BAO measurements, we use the Markov Chain Monte Carlo (MCMC) method to determine the cosmological parameters in various extensions to the standard $\Lambda$CDM models, such as the equation of state (EoS) of dark energy $w$, the curvature of the universe $\Omega_k$, the total neutrino mass $\sum{m_\nu}$, the effective number of neutrinos $N_{\rm eff}$, and the parameters associated with the power spectrum of primordial fluctuations. Our paper is organized as follows: In section \ref{Method} we describe the method and the observational data sets used in our calculations; section \ref{results} contains our numerical constraints on various cosmological models from the $H(z)$ data. The last section \ref{summary} is dedicated to the conclusions.

\section{Method and Data}\label{Method}


We perform a global analysis by employing the publicly available MCMC package {\tt CosmoMC} \cite{cosmomc}. Assuming the purely adiabatic initial conditions, we use the current observations to constrain cosmological parameters of several extensions to the standard $\Lambda$CDM models, which have been summarized in Table \ref{table-parameters}.

\begin{table*}[t]
\caption{Cosmological parameters used in our analyses. The block above the middle line shows the basic six parameters in the standard $\Lambda$CDM model, and the block below the line includes the derived parameters in each extended model.} \label{table-parameters}
\begin{center}
\renewcommand{\arraystretch}{1.5}
\begin{tabularx}{110mm}{lXl}
\hline
\hline
Parameter &    Description              & Prior range    \\
\hline
$\Omega_b h^2$ &   Physical baryon density today&    $[0.005,0.1]$     \\
$\Omega_{dm} h^2$  &  Physical dark matter density today  & $[0.01,0.99]$        \\
$\Theta$          &  100 times angular size of sound horizon&$[0.5,10]$  \\
$\tau$            & Re-ionization optical depth   &$[0.01,0.8]$ \\
$n_s$     &   Scalar spectral index at $k_{s0} = 0.002$ ${\rm Mpc}^{-1}$ & $[0.5,1.5]$  \\
$\ln{(10^{10}A_s)}$ &      Amplitude of the primordial curvature perturbations at $k_{s0} = 0.002$ ${\rm Mpc}^{-1}$                   &$[2.7,4.0]$ \\
\hline
$N_{\rm{eff}}$        &  Effective number of neutrinos& $[1.0,10.0]$ \\
$f_{\nu}$       & Fraction of the dark matter in the form of massive neutrinos   &$[0,0.5]$\\
$\Omega_k$    &  Spatial curvature parameter today     & $[-0.3,0.3]$\\
$\alpha_s$  & Running of the spectral index & $[-0.3,0.3]$ \\
$r$            &  The tensor to scalar ratio of the primordial spectrum  & $[0,2]$       \\
$w$             & Constant dark energy equation of state  &$[-5.0,3.0]$ \\
\hline\hline
\end{tabularx}
\end{center}
\end{table*}


In our analysis, we consider the following cosmological probes: i) power spectra of CMB temperature and polarization anisotropies; ii) the baryon acoustic oscillation in the galaxy power spectra; iii) luminosity distances of type Ia supernovae; iv) measurements of the Hubble parameter.

To incorporate the WMAP9 CMB temperature and polarization power spectra, we use the routines for computing the likelihood supplied by the WMAP team \cite{wmap}.  The WMAP9 polarization data are composed of TE/EE/BB power spectra on large scales ($2 \leq \ell \leq 23$) and TE power spectra on small scales ($24 \leq \ell \leq 800$), while the WMAP9 temperature data includes the CMB anisotropies on scales $2 \leq \ell \leq 1200$.

Baryon Acoustic Oscillations provides an efficient method for measuring the expansion history by using features in the clustering of galaxies within large scale surveys as a ruler with which to measure the distance-redshift relation. Since the current BAO data are not accurate enough, one can only determine an effective distance \cite{baosdss}:
\begin{equation}
D_V(z)=[(1+z)^2D_A^2(z)cz/H(z)]^{1/3}~.
\end{equation}
In this paper we use the BAO measurement from the 6dF Galaxy Redshift Survey (6dFGRS) at a low redshift ($z = 0.106$) \cite{6dfgrs}, the measurement of the BAO scale based on a re-analysis of the Luminous Red Galaxies (LRG) sample from Sloan Digital Sky Survey (SDSS) Data Release 7 at the median redshift ($z = 0.35$) \cite{sdssdr7}, the BAO signal from BOSS CMASS DR9 data at redshift ($z = 0.57$) \cite{sdssdr9}, the BAO measurement from the WiggleZ survey at $z=0.44$, $z=0.60$ and $z=0.73$ \cite{wigglez}, and the latest measurement of BAO at high redshift of $z=2.3$ from the analysis of Ly-$\alpha$ forest of BOSS quasars \cite{busca}.

In this paper we use the latest SNIa data sets from the Supernova Cosmology Project, ``Union Compilation 2.1'', which consists of 580 samples and spans the redshift range $0\lesssim{z}\lesssim1.55$ \cite{union2}. This data set also provides the covariance matrix of data with and without systematic errors. In order to be conservative, we use the covariance matrix with systematic errors. When calculating the likelihood from SNIa, we marginalize over the absolute magnitude $M$, which is a nuisance parameter, as done in refs. \cite{SNMethod1,SNMethod2}.

\begin{table*}[t] 
\caption{$H(z)$ measurements and their errors in units of ${\rm km\,s^{-1}\,Mpc^{-1}}$. (${\tt Reference. -}$ [1] Gazta\~naga {\it et al.} (2009); [2] Stern {\it et al.} (2010); [3] Moresco {\it et al.} (2012); [4] Zhang {\it et al.} (2012); [5] Simon {\it et al.} (2005).)}\label{Hz}
\begin{center}
\begin{tabular}{c|c|c|c|c|c|c|c|c|c|c|c|c}

\hline
\hline

$z$ & $~0.07~$ & $~0.09~$ & $~0.12~$ & $~0.17~$ & $0.1791$ & $0.1993$ & $~0.2~~$ & $~0.24~$ & $~0.27~$ & $~0.28~$ & $0.3519$ & $~0.40~$ \\

$H(z)$ & $69$ & $69$ & $68.6$ & $83$ & $75$ & $75$ & $72.9$ & $79.69$ & $77$ & $88.8$ & $83$ & $95$ \\

$\sigma_{H(z)}$ & $19.6$ & $12$ & $26.2$ & $8$ & $4$ & $5$ & $29.6$ & $2.65$ & $14$ & $36.6$ & $14$ & $17$ \\

Ref. & $[4]$ & $[2]$ & $[4]$ & $[2]$ & $[3]$ & $[3]$ & $[4]$ & $[1]$ & $[2]$ & $[4]$ & $[3]$ & $[2]$ \\

\hline\hline

$~0.43~$ & $0.48$ & $0.5929$ & $0.6797$ & $0.7812$ & $0.8754$ & $0.88$ & $0.9$ & $1.037$ & $1.3$ & $1.43$ & $1.53$ & $1.75$\\

$86.45$ & $97$ & $104$ & $92$ & $105$ & $125$ & $90$ & $117$ & $154$ & $168$ & $177$ & $140$ & $202$\\

$3.68$ & $62$ & $13$ & $8$ & $12$ & $17$ & $40$ & $23$ & $20$ & $17$ & $18$ & $14$ & $40$\\

$[1]$ & $[2]$ & $[3]$ & $[3]$ & $[3]$ & $[3]$ & $[2]$ & $[2]$ & $[3]$ & $[5]$ & $[5]$ & $[5]$ & $[5]$\\

\hline\hline
\end{tabular}
\end{center} 
\end{table*}

Finally, we include the two kinds of $H(z)$ data in our analyses: a) ``$Hz~data$'': the direct measurements on the Hubble parameter at high redshifts $H(z)$. Here, we adopt 25 Hubble parameter data obtained from refs. \cite{gaztanaga,stern,moresco,simon,zhang2012} which are listed in Table \ref{Hz}. b) ``HST $H_0$ prior'': the HST measurement on the Hubble constant, $H_0=73.8\pm{2.4}$ ${\rm km\,s^{-1}\,Mpc^{-1}}$ ($68\%$ C.L.), which is obtained from the magnitude-redshift relation of 253 low-z Type Ia supernovae at $z < 0.1$ by the Wide Field Camera 3 (WFC3) on the Hubble Space Telescope (HST). We calculate the $\chi^2$ value of the $H(z)$ data by using
\begin{equation}
\chi^2_{\rm H(z)}=\sum_i\frac{(H^{\rm th}(z_i)-H^{\rm obs}(z_i))^2}{\sigma^2_{\rm H}(z_i)}~,
\end{equation}
where $H^{\rm th}(z)$ and $H^{\rm obs}(z)$ are the theoretical and observed values of Hubble parameter at redshift $z$, and $\sigma_{\rm H}$ denotes the error bar of observed $H(z)$ data.

\section{Numerical Results}\label{results}

In this section we present our global fitting results of the cosmological parameters determined from the latest observational data and mainly focus on the effect of constraints from the $H(z)$ data. Here, we consider three basic data combinations: WMAP9, WMAP9+SNIa and WMAP9+BAO, and find that the effects of additional $H(z)$ data on cosmological parameters in these three cases are almost identical. Therefore, in the following analyses, we mainly use the WMAP9+SNIa combination as an example to show the effect of $H(z)$ data and list the numerical results of these three cases in tables.

\subsection{Effective Number of Neutrinos}\label{nnueff}
Neutrinos fix the expansion rate during the cosmological era when the Universe is dominated by radiation. Their contribution to the total radiation content can be parameterized in terms of the effective number of neutrino, $N_{\rm eff}$, which is directly related to the expansion history of universe at early time. In the standard cosmology, based on the analysis of neutrino decoupling, three active neutrinos contribute as $N_{\rm eff}=3.046$. Any departure from this value would be due to non-standard neutrino features, such as the sterile neutrinos \cite{cho,ciuffoli:2012yd}. In this subsection, we consider the constraints on the effective number of neutrino, especially from the $H(z)$ data. We summarize the numerical results from different data combinations in Table \ref{neff}.

\begin{table*}[t] 
\caption{$68\%$ constraints on the effective number of neutrino $N_{\rm eff}$ and some other cosmological parameters from different data combinations.}\label{neff}
\begin{center}
\begin{tabular}{c|c |c| c|c |c |c|c| c| c}
\hline
& \multicolumn{3}{c|}{WMAP9}& \multicolumn{3}{c|}{WMAP9+SNIa}& \multicolumn{3}{c}{WMAP9+BAO} \\
\cline{2-10}
& $N_{\rm eff}$ & $H_0$ & $\Omega_m$ & $N_{\rm eff}$ & $H_0$ & $\Omega_m$ &$N_{\rm eff}$ & $H_0$ & $\Omega_m$\\
\hline
+&$4.18^{+2.06}_{-1.98}$&$74.8^{+9.0}_{-8.8}$&$0.278^{+0.025}_{-0.026}$&$4.21^{+1.95}_{-1.88}$&$75.3^{+8.4}_{-8.1}$&$0.273\pm{0.022}$&$2.97^{+0.69}_{-0.70}$&$68.3\pm{3.1}$&$0.296\pm{0.011}$ \\
\hline
+ $Hz~data$&$3.21^{+0.38}_{-0.39}$&$70.6\pm{1.6}$&$0.282\pm{0.022}$&$3.18^{+0.38}_{-0.37}$&$70.6\pm{1.4}$&$0.279^{+0.020}_{-0.019}$&$3.29\pm{0.31}$&$69.8\pm{1.3}$&$0.295\pm{0.011}$\\
\hline
+ HST $H_0$&$3.80^{+0.76}_{-0.74}$&$73.7\pm{2.4}$&$0.275^{+0.025}_{-0.026}$&$3.82^{+0.72}_{-0.71}$&$73.8\pm{2.4}$&$0.274^{+0.020}_{-0.022}$&$3.77^{+0.49}_{-0.50}$&$71.9\pm{2.0}$&$0.292^{+0.011}_{-0.010}$\\
\hline
\end{tabular}
\end{center} 
\end{table*}

\begin{figure}[t]
\begin{center}
\includegraphics[scale=0.35]{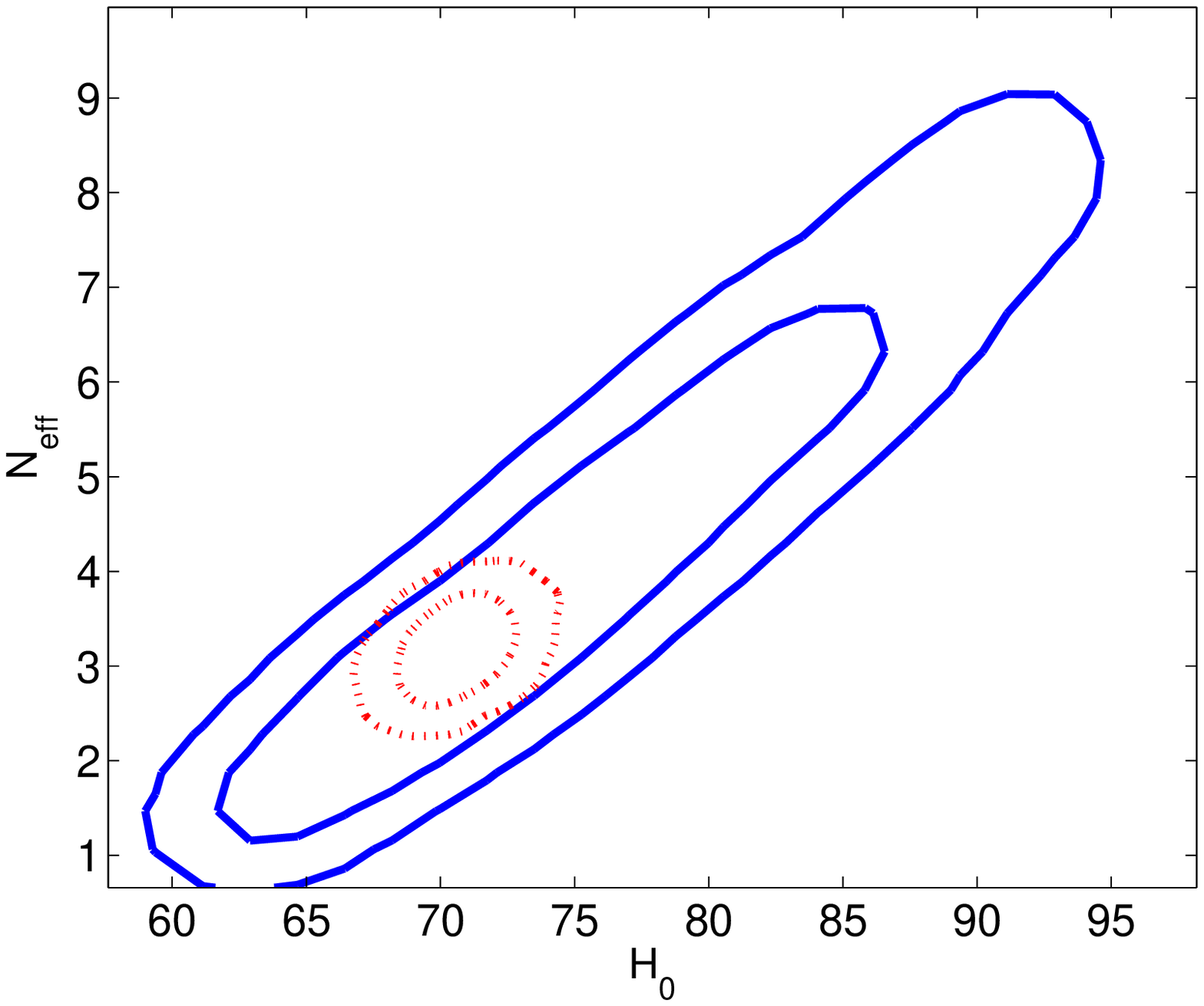}
\includegraphics[scale=0.35]{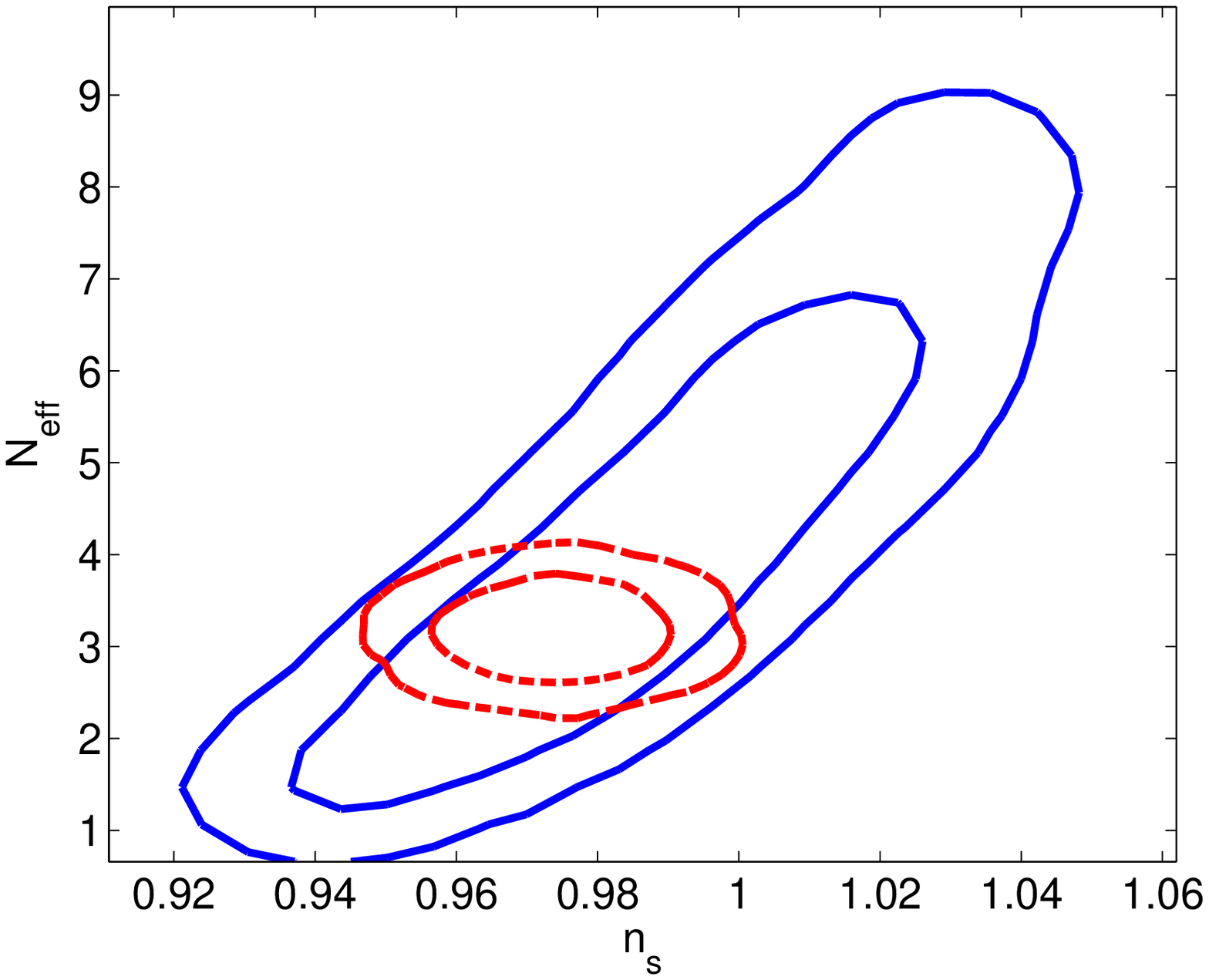}
\caption{Two-dimensional marginalized distribution between $N_{\rm eff}$ and $H_0$ (left panel), $n_s$ (right panel). The blue solid and red dotted lines are obtained from WMAP9+SNIa and WMAP9+SNIa+$Hz$, respectively.}\label{fig_neff_hz}
\end{center}
\end{figure}

Since $N_{\rm eff}$ can be written in terms of $\Omega_mh^2$ and the redshift of matter-radiation equality, $z_{\rm eq}$, there are strong degeneracies present between $N_{\rm eff}$, $\Omega_mh^2$ and the Hubble parameter $H_0$ \cite{wmap_neff,verde_neff}. Constraints on $N_{\rm eff}$ can thus be strongly improved by combining with measurements of Hubble parameter. In Table \ref{neff} we find that WMAP9+SNIa data can only give very weak constraint on $N_{\rm eff}$. Adding the ``$Hz~data$''  significantly improves the $68\%$ C.L. constraint,
\begin{equation}
N_{\rm eff}=3.18^{+0.38}_{-0.37}~.
\end{equation}
The standard value of $N_{\rm eff} = 3.046$ remains well within the $68\%$ confidence intervals, which is consistent with previous results \cite{wmap,planck_fit,xia_neff}. We also show the two-dimensional contour between $N_{\rm eff}$ and $H_0$ in the left panel of Figure \ref{fig_neff_hz}, which is clearly shown that $N_{\rm eff}$ is strongly correlated with $H_0$. Taking the $Hz~data$ into account could shrink the $1\,\sigma$ error bar of $N_{\rm eff}$ significantly, due to the better constraint on the current Hubble constant $H_0$.

\begin{figure}[t]
\begin{center}
\includegraphics[scale=0.35]{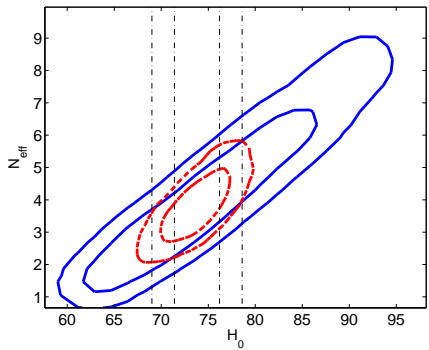}
\caption{Two-dimensional marginalized distribution between $N_{\rm eff}$ and $H_0$. The blue solid lines are obtained from WMAP9+SNIa, and the red dotted lines denotes the constraints from WMAP9+SNIa+HST $H_0$. Four vertical dashed lines denote the $1,2\,\sigma$ limits of the $H_0$ prior from HST.}\label{fig_neff_h0}
\end{center}
\end{figure}

More interestingly, we find that the high redshift $Hz~data$ is helpful to break the degeneracy between $N_{\rm eff}$ and $n_s$. Using WMAP9+SNIa, we find the $68\%$ C.L. limit of $n_s=0.974 \pm 0.013$ in the standard $\Lambda$CDM model. When we vary $N_{\rm eff}$ in our analysis, the constraint on $n_s$ is weakened by a factor of two, namely $n_s=0.987^{+0.028}_{-0.027}$ ($68\%$ C.L.). The coefficient of the correlation between the $N_{\rm eff}$ and $n_s$ is $cov(N_{\rm eff},n_s)=0.88$. Due to this strongly correlation, shown in the right panel of Figure \ref{fig_neff_hz}, including $N_{\rm eff}$ into the cosmological model would weaken the constraint on $n_s$ \cite{ns_neff}. When we add the ``$Hz~data$'', the constraint on $n_s$ becomes tighter, $n_s=0.974\pm{0.011}$ at $68\%$ confidence level. The correlation between $N_{\rm eff}$ and $n_s$ has been totally broken, namely the coefficient of the correlation $cov(N_{\rm eff},n_s)=0.02$.

Now we consider the effects of the HST $H_0$ prior in the calculations. As we discuss above, $N_{\rm eff}$ is strongly correlated with $H_0$. The larger $H_0$ is, the larger $N_{\rm eff}$ the observational data favor. So, when we use the HST prior which gives a slightly larger value of $H_0$, the obtained median value of $N_{\rm eff}$ becomes larger, consequently. The $68\%$ C.L. constraint is
\begin{equation}
N_{\rm eff}=3.82^{+0.72}_{-0.71}~,
\end{equation}
which displays a slight preference for an extra relativistic relic.

We also show the effect of the HST $H_0$ prior in the contour of $(N_{\rm eff},H_0)$ (Figure \ref{fig_neff_h0}). When we use WMAP9+SNIa data alone, the allowed ranges of $H_0$ and $N_{\rm eff}$ are very large (blue solid lines). When adding a strong prior of $H_0$ into the analysis (vertical dashed lines), the joint two-dimensional contours shrink to the overlapped area between these two data sets and are highly dependent on the median value of the $H_0$ prior.

\subsection{Total Neutrino Mass}
\begin{table*}[t] 
\caption{$68\%$ constraints on the total neutrino mass $\sum{m_\nu}$ and some other cosmological parameters from different data combinations. For the total neutrino mass, we quote the $95\%$ upper limits instead.}\label{mnu}
\begin{center}
\begin{tabular}{c|c |c| c|c |c |c|c| c| c}
\hline
& \multicolumn{3}{c|}{WMAP9}& \multicolumn{3}{c|}{WMAP9+SNIa}& \multicolumn{3}{c}{WMAP9+BAO} \\
\cline{2-10}
& $\sum{m_\nu}$ & $H_0$ & $\Omega_m$ & $\sum{m_\nu}$ & $H_0$ & $\Omega_m$ &$\sum{m_\nu}$ & $H_0$ & $\Omega_m$\\
\hline
+&$<1.2$eV&$65.5^{+4.4}_{-4.7}$&$0.338^{+0.063}_{-0.059}$&$<0.62$eV&$68.6\pm{2.4}$&$0.294\pm{0.028}$&$<0.59$eV&$67.8\pm{1.1}$&$0.303\pm{0.012}$ \\
\hline
+$Hz~data$&$<0.46$eV&$69.5\pm{1.7}$&$0.285\pm{0.020}$&$<0.44$eV&$69.7\pm{1.6}$&$0.281\pm{0.018}$&$<0.49$eV&$68.41^{+0.91}_{-0.89}$&$0.297\pm{0.011}$\\
\hline
+HST $H_0$&$<0.38$eV&$71.3\pm{1.8}$&$0.264\pm{0.019}$&$<0.40$eV&$71.0\pm{1.7}$&$0.268^{+0.018}_{-0.017}$&$<0.47$eV&$68.80^{+0.96}_{-0.98}$&$0.293\pm{0.011}$\\
\hline
\end{tabular}
\end{center} 
\end{table*}

Detecting the neutrino mass is one of the challenges of modern physics. Currently the neutrino oscillation experiments, such as atmospheric neutrinos experiments \cite{atmospheric1,atmospheric2,atmospheric3,atmospheric4,atmospheric5}, solar neutrinos experiments \cite{solar1,solar2,solar3,solar4,solar5,solar6,solar7}, reactor neutrinos experiments \cite{reactor1,reactor2} and accelerator beam neutrinos experiments \cite{accelerator1,accelerator2}, have confirmed that the neutrinos are massive, but give no hint on their absolute mass scale. Fortunately, cosmological observational data can provide the crucial complementary information on absolute neutrino masses, because massive neutrinos leave imprints on the cosmological observations, such as the Hubble diagram, CMB temperature power spectrum, and LSS matter power spectrum \cite{mnu_rev}. In this subsection, we consider the constraints on the total neutrino mass from the $H(z)$ data. Assuming $N_{\rm eff}=3.046$, we summarize the numerical results from different data combinations in Table \ref{mnu}.

\begin{figure}[t]
\begin{center}
\includegraphics[scale=0.27]{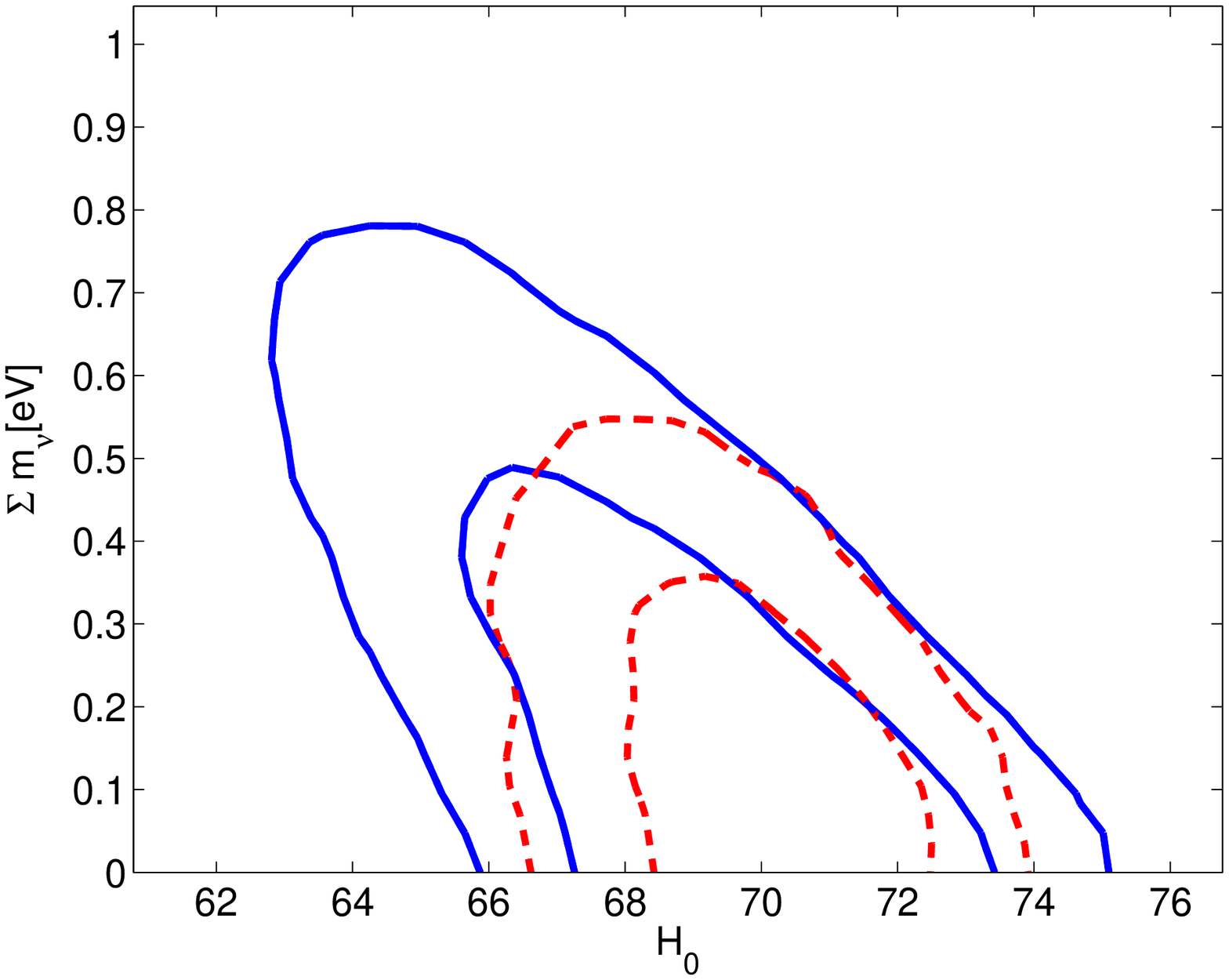}
\includegraphics[scale=0.26]{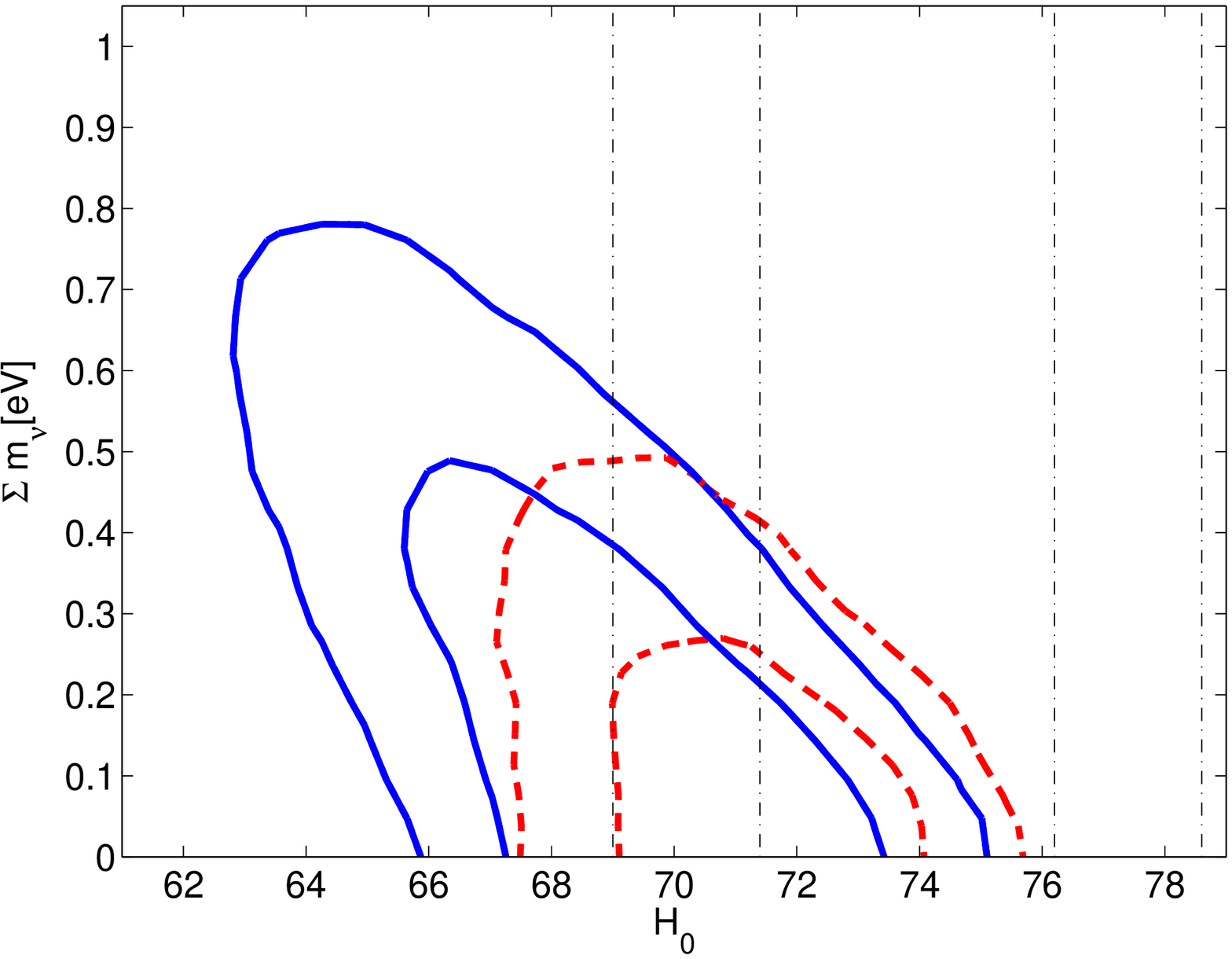}
\caption{Two-dimensional marginalized distribution between $\sum m_{\nu}$ and $H_0$. The blue solid lines are obtained from WMAP9+SNIa, and the red dotted lines denotes the constraints from WMAP9+SNIa+$Hz~data$ (left panel) and WMAP9+SNIa+HST $H_0$ (right panel). Four vertical dashed lines denote the $1,2\,\sigma$ limits of the $H_0$ prior from HST.}\label{fig_mnu_2d}
\end{center}
\end{figure}

Due to the strong degeneracies present between cosmological parameters, primary CMB anisotropies alone can place only weak constraints on the total neutrino mass. In the flat $\Lambda$CDM framework, the WMAP9 data alone weakly constrain the total neutrino mass, $\sum m_{\nu}<1.2$eV ($95\%$ C.L.). Adding the ``Union2.1'' data breaks these degeneracies and the constraint on the total neutrino mass significantly improves to $\sum m_{\nu}<0.62$eV ($95\%$ C.L.) for the combined WMAP9 and SNIa datasets. The efficient way to improve the limit is to introduce observations that constrain the Hubble constant. When we include the ``$Hz~data$'' into our calculations, the constraint on $\sum m_{\nu}$ becomes tighter further,
\begin{equation}
\sum m_{\nu}<0.44~{\rm eV}~~(95\%~{\rm C.L.})~.
\end{equation}
In Figure \ref{fig_mnu_2d} we show the two-dimensional contours in the $(\sum m_{\nu},H_0)$ panel from different data combinations. $\sum m_{\nu}$ and $H_0$ are anti-correlated, since the most prominent effect caused by light neutrino is to change the position of the first peak and it is absorbed into a lowering shift of the Hubble constant \cite{mnu_effect}. In the left panel, we can see that adding the ``$Hz~data$" improves the constraints on the current Hubble constant $H_0$ and reduces the correlations between $\sum m_{\nu}$ and $H_0$. Therefore, we obtain tighter constraint on $\sum m_{\nu}$ than WMAP9+SNIa data.

We now present the constraints on $\sum m_{\nu}$ obtained from the direct HST $H_0$ prior. In the right panel of Figure \ref{fig_mnu_2d}, we show the two-dimensional constraints on $\sum m_{\nu}$ from WMAP9+SNIa+HST $H_0$. As can be seen, the WMAP9+SNIa data give the $68\%$ constraint $H_0=68.6\pm{2.4}$ ${\rm km\,s^{-1}\,Mpc^{-1}}$, which is lower than the HST prior $H_0=73.8$ ${\rm km\,s^{-1}\,Mpc^{-1}}$. Adding the HST $H_0$ prior would shift the limit of $H_0$ towards a higher value, $H_0=71.0\pm{1.7}$ ${\rm km\,s^{-1}\,Mpc^{-1}}$ ($68\%$ C.L.). Due to the anti-correlation between $\sum m_{\nu}$ and $H_0$, the upper limit of $\sum m_{\nu}$ is strongly suppressed,
\begin{equation}
\sum m_{\nu} < 0.40~{\rm eV}~~(95\%~{\rm C.L.})~.
\end{equation}

\subsection{Freeing both Effective Number of Neutrinos and Total Neutrino Mass simultaneously}

\begin{table*}[t] 
\caption{$68\%$ constraints on the effective number of neutrino $N_{\rm eff}$, the total neutrino mass $\sum{m_\nu}$ and some other cosmological parameters from different data combinations. For the total neutrino mass, we quote the $95\%$ upper limits instead.}\label{neff and mnu}
\begin{center}
\begin{tabular}{c|c |c| c|c |c |c|c| c}
\hline
& \multicolumn{4}{c|}{WMAP9}& \multicolumn{4}{c}{WMAP9+SNIa}\\
\cline{2-9}
& $N_{\rm eff}$ & $\sum{m_\nu}$ &$H_0$ & $\Omega_m$ & $N_{\rm eff}$ &$\sum{m_\nu}$& $H_0$ & $\Omega_m$ \\
\hline
+&$4.49^{+2.02}_{-1.97}$&$<1.6$eV&$70.5^{+8.6}_{-8.5}$&$0.344^{+0.068}_{-0.060}$&$4.3\pm{1.9}$&$<0.78$eV&$73.3^{+8.4}_{-7.9}$&$0.296^{+0.028}_{-0.027}$\\
\hline
+ $Hz~data$&$3.68^{+0.55}_{-0.54}$&$<1.1$eV&$69.0\pm{2.0}$&$0.320^{+0.042}_{-0.040}$&$3.39\pm{0.42}$&$<0.68$eV&$69.9\pm{1.6}$&$0.295\pm{0.023}$\\
\hline
+ HST $H_0$&$5.0^{+1.4}_{-1.2}$&$<1.6$eV&$73.4\pm{2.2}$&$0.328^{+0.057}_{-0.051}$&$4.22\pm{0.81}$&$<0.77$eV&$73.6\pm{2.3}$&$0.291^{+0.026}_{-0.027}$\\
\hline

\end{tabular}
\end{center} 

\begin{center}
\begin{tabular}{c|c |c| c|c}
\hline
& \multicolumn{4}{c}{WMAP9+BAO} \\
\cline{2-5}
& $N_{\rm eff}$ & $\sum{m_\nu}$ &$H_0$ & $\Omega_m$ \\
\hline
+&$3.04^{+0.71}_{-0.70}$&$<0.60$eV&$67.7^{+3.2}_{-3.0}$&$0.303\pm{0.012}$ \\
\hline
+ $Hz~data$&$3.41^{+0.33}_{-0.32}$&$<0.63$eV&$69.4\pm{1.3}$&$0.301\pm{0.011}$\\
\hline
+ HST $H_0$&$3.90^{+0.52}_{-0.51}$&$<0.62$eV&$71.8\pm{2.0}$&$0.298\pm{0.011}$\\
\hline

\end{tabular}
\end{center} 
\end{table*}

We consider varying both of the effective number of neutrinos $N_{\rm eff}$ and total neutrino mass $\sum m_{\nu}$ simultaneously in the fitting, and discuss effects from the degeneracy between them in this subsection.

Firstly, varying the two parameters in the same time will enlarge the constraining error of cosmological parameters, since it adds new degree of freedom in the fitting. The detailed numerical results of this fitting from different data combinations are summarized in Table \ref{neff and mnu}.

As can be seen in the previous subsection that, $N_{\rm eff}$ is positively correlated with $H_0$, while $\sum m_{\nu}$ is negatively correlated with $H_0$, varying both of $N_{\rm eff}$ and $\sum m_{\nu}$ in the same time during the fitting can cancel the effects from these two kinds of degeneracies in some sense, however, the effect is not obvious from seeing error bars listed in the table. Using the data combination of WMAP9$+$SNIa or WMAP9$+$BAO data sets, the results shows that the constraints on both of $N_{\rm eff}$ and $\sum m_{\nu}$ are weaker than those from freeing $N_{\rm eff}$ or $\sum m_{\nu}$ only. When freeing both of $N_{\rm eff}$ and $\sum m_{\nu}$ simultaneously, the effects from the degeneracy between $N_{\rm eff}$ and other cosmological parameters are dominant, and it is can be seen by observing the detailed numerical results. With the data of WMAP9+SNIa+$Hz~data$, we obtain the $68\%$ constraints on $N_{\rm eff}$ and $H_0$ are $3.39\pm{0.42}$ and $69.9\pm{1.6}$ respectively, which is very consistent with the results of subsection \ref{nnueff} which frees $N_{\rm eff}$ and fixes $\sum m_{\nu}$ during the fitting procedure. The same effect can also be reflected when we use the WMAP9+SNIa+HST $H_0$ data.

$N_{\rm eff}$ and $\sum m_{\nu}$ are positively correlated with each other. Using the WMAP9+SNIa data, the coefficient of the correlation between the two is $cov(N_{\rm eff}, \sum m_{\nu})=0.27$. When adding the $``Hz~data"$ and HST $H_0$ prior, the constraints becomes tighter and the coefficient of the correlation between them slightly increase to $cov(N_{\rm eff}, \sum m_{\nu})=0.55$ and $cov(N_{\rm eff}, \sum m_{\nu})=0.57$, respectively. In the left and right panels of Figure 4, we show the two-dimensional constraints on $\sum m_{\nu}$ and $N_{\rm eff}$ from WMAP9+SNIa+HST $H_0$ and WMAP9+SNIa+$Hz~data$. As an example of such positive degeneracy, when comparing with the results from WMAP9+SNIa+HST $H_0$, a more tight constraint on $\sum m_{\nu}$ brought by WMAP9+SNIa+$Hz~data$ prefers to a lower value of $N_{\rm eff}$, which can be seen in Table \ref{neff and mnu}.

\begin{figure}[t]
\begin{center}
\includegraphics[scale=0.30]{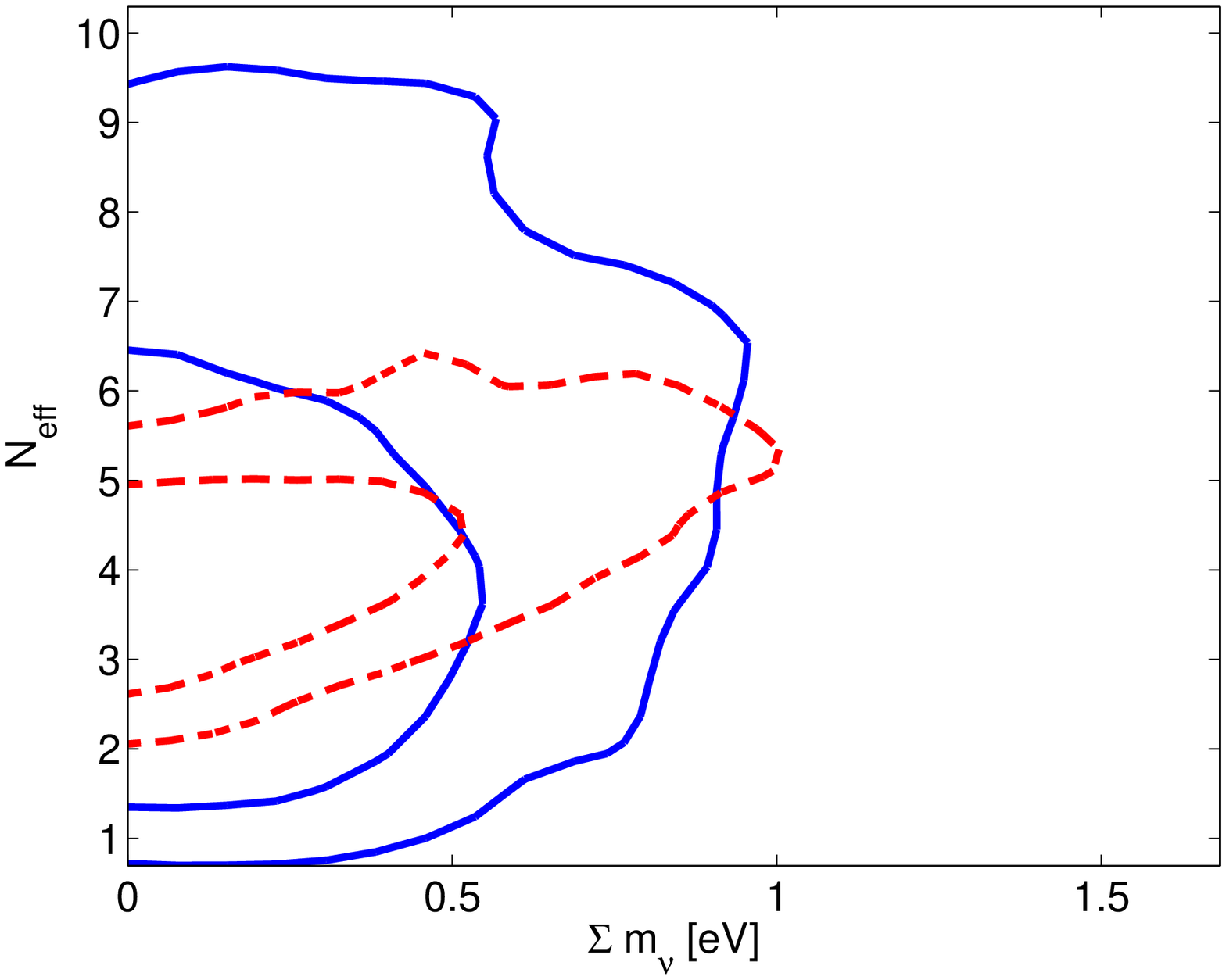}
\includegraphics[scale=0.30]{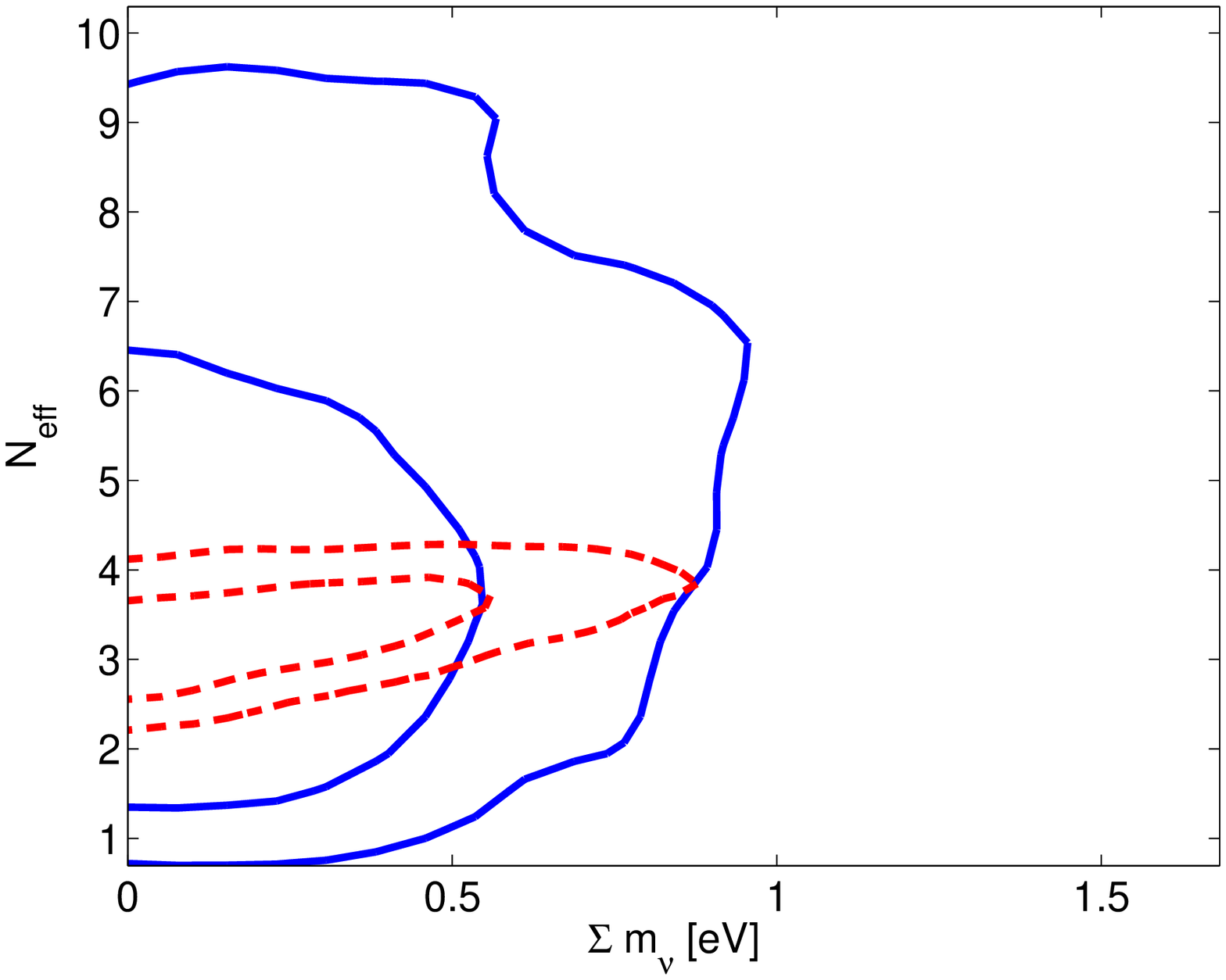}
\caption{Two-dimensional marginalized distribution between $\sum m_{\nu}$ and $N_{\rm eff}$. The blue solid lines are obtained from WMAP9+SNIa, and the red dotted lines denotes the constraints from WMAP9+SNIa+HST $H_0$ (left panel) and WMAP9+SNIa+$Hz~data$ (right panel), respectively.}\label{fig_mnu_nnu_2d}
\end{center}
\end{figure}

\subsection{Spatial Curvature}

\begin{table*}[t] 
\caption{$68\%$ constraints on the curvature $\Omega_k$ and some other cosmological parameters from different data combinations. }\label{omk}
\begin{center}
\begin{tabular}{c|c |c| c|c |c |c|c| c| c}
\hline
& \multicolumn{3}{c|}{WMAP9}& \multicolumn{3}{c|}{WMAP9+SNIa}& \multicolumn{3}{c}{WMAP9+BAO} \\
\cline{2-10}
& $100\Omega_k$ & $H_0$ & $\Omega_m$ & $100\Omega_k$ & $H_0$ & $\Omega_m$ &$100\Omega_k$ & $H_0$ & $\Omega_m$\\
\hline
+&$-1.8^{+2.8}_{-2.9}$&$66\pm{12}$&$0.35\pm{0.13}$&$-0.19^{+0.99}_{-1.00}$&$70.2^{+4.9}_{-5.0}$&$0.280\pm{0.040}$&$-0.46\pm{0.44}$&$68.1\pm{1.2}$&$0.294\pm{0.012}$ \\
\hline
+$Hz~data$&$0.02\pm{0.44}$&$70.3^{+1.7}_{-1.8}$&$0.278\pm{0.018}$&$0.02^{+0.42}_{-0.43}$&$70.5^{+1.7}_{-1.6}$&$0.276\pm{0.017}$&$-0.22\pm{0.39}$&$68.72^{+0.89}_{-0.92}$&$0.292\pm{0.011}$\\
\hline
+HST $H_0$&$0.39^{+0.50}_{-0.51}$&$73.2^{+2.4}_{-2.3}$&$0.254\pm{0.019}$&$0.41^{+0.47}_{-0.48}$&$73.1\pm{2.2}$&$0.255\pm{0.018}$&$-0.27^{+0.41}_{-0.42}$&$69.07\pm{0.97}$&$0.287\pm{0.011}$\\
\hline

\end{tabular}
\end{center} 
\end{table*}

Inflationary models predict that our Universe should be very accurately spatially flat. Observational limits on spatial curvature therefore offer important additional constraints on inflationary models and fundamental physics. In this subsection, we discuss the constraints on $\Omega_k$ when taking the Hubble parameters data into account and summarize the numerical results from different data combinations in Table \ref{omk}.

Using the WMAP9 data alone, the curvature can be constrained very weakly, $\Omega_k=-0.018^{+0.028}_{-0.029}$ (68$\%$ C.L.), due to the well-known geometric degeneracy on the CMB temperature power spectrum \cite{bond, zaldarriaga}. This is a near perfect degeneracy can be broken with the addition of probes of late time physics, including BAO, SNIa, and measurement of the Hubble constant \cite{wmap3}. When we combine the WMAP9 and SNIa data together, the constraint on the curvature becomes tighter, $\Omega_k=-0.0019^{+0.0099}_{-0.0100}$ at $68\%$ confidence level. If we use the ``$Hz~data$'' instead of the SNIa data, the error bar of $\Omega_k$ is shrunk significantly, namely the $68\%$ constraint $\Omega_k=0.0002\pm{0.0044}$. The ``$Hz~data$'' is very helpful to break this geometric degeneracy. When combining WMAP9, SNIa and ``$Hz~data$'' together, we obtain very tight constraint on the curvature
\begin{equation}
\Omega_k = 0.0002^{+0.0042}_{-0.0043}~~(68\%~{\rm C.L.})~,
\end{equation}
which suggest that our Universe is spatially flat. These limits are consistent with the results reported by the WMAP9 \cite{wmap} and Planck data \cite{planck_fit}.

Besides the ``$Hz~data$'', the direct $H_0$ prior could also affect the constraint of $\Omega_k$ and break the degeneracy. When using the HST $H_0$ prior, we obtain a slightly large median value of the curvature, $\Omega_k=0.0041^{+0.0047}_{-0.0048}$. However, the flat universe remains well within the $68\%$ confidence intervals. There is no evidence from the current observational data for any departure from a spatially flat geometry.

\subsection{Dark Energy Equation of State}

\begin{table*}[t] 
\caption{$68\%$ constraints on the constant EoS of dark energy $w$ and some other cosmological parameters from different data combinations. }\label{wde}
\begin{center}\label{w}
\begin{tabular}{c|c |c| c|c |c |c|c| c| c}
\hline
& \multicolumn{3}{c|}{WMAP9}& \multicolumn{3}{c|}{WMAP9+SNIa}& \multicolumn{3}{c}{WMAP9+BAO} \\
\cline{2-10}
& $w$ & $H_0$ & $\Omega_m$ & $w$ & $H_0$ & $\Omega_m$ &$w$ & $H_0$ & $\Omega_m$\\
\hline
+&$-$&$-$&$-$&$-1.011^{+0.074}_{-0.073}$&$70.6^{+2.3}_{-2.2}$&$0.274\pm{0.021}$&$-0.96\pm{0.15}$&$67.9\pm{3.1}$&$0.301\pm{0.021}$ \\
\hline
+$Hz~data$&$-0.98\pm{0.14}$&$69.8^{+3.7}_{-3.5}$&$0.282^{+0.028}_{-0.029}$&$-0.997\pm{0.072}$&$70.3\pm{1.8}$&$0.277\pm{0.016}$&$-0.98^{+0.12}_{-0.13}$&$68.4^{+2.4}_{-2.3}$&$0.297^{+0.017}_{-0.016}$\\
\hline
+HST $H_0~$&$-1.095^{+0.097}_{-0.096}$&$73.6\pm{2.3}$&$0.252\pm{0.018}$&$-1.038^{+0.070}_{-0.071}$&$72.1^{+1.7}_{-1.6}$&$0.262\pm{0.016}$&$-1.14\pm{0.11}$&$71.8^{+2.1}_{-2.2}$&$0.276\pm{0.014}$\\
\hline
\end{tabular}
\end{center} 
\end{table*}

A major challenge for cosmology is to elucidate the nature of the dark energy driving the accelerated expansion of the Universe. A cosmological constant, the simplest candidate of dark energy where the EoS $w\equiv -1$, suffers from the well-known fine-tuning and coincidence problems \cite{lcdm1,lcdm2,lcdm3}. Alternatively, dynamical dark energy models with the rolling scalar fields have been proposed, such as quintessence \cite{lcdm2,lcdm3,quintessence3,quintessence4}, the ghost field of phantom \cite{phantom}, the model of $k$ essence which has a noncanonical kinetic term \cite{kessence1,kessence2,kessence3} and the quintom model \cite{quintom1,quintom2,quintom3,quintom4,pr_rev}. Assuming the flat universe, in this subsection we study the dark energy model with a constant equation of state $w$ from the current data, especially from the $H(z)$ data. Since the current data are not accurate enough, here we do not consider the dark energy model with a time-evolving EoS. In Table \ref{wde} we summarize the numerical results on some parameters from different data combinations.

\begin{figure}[t]
\begin{center}
\includegraphics[scale=0.31]{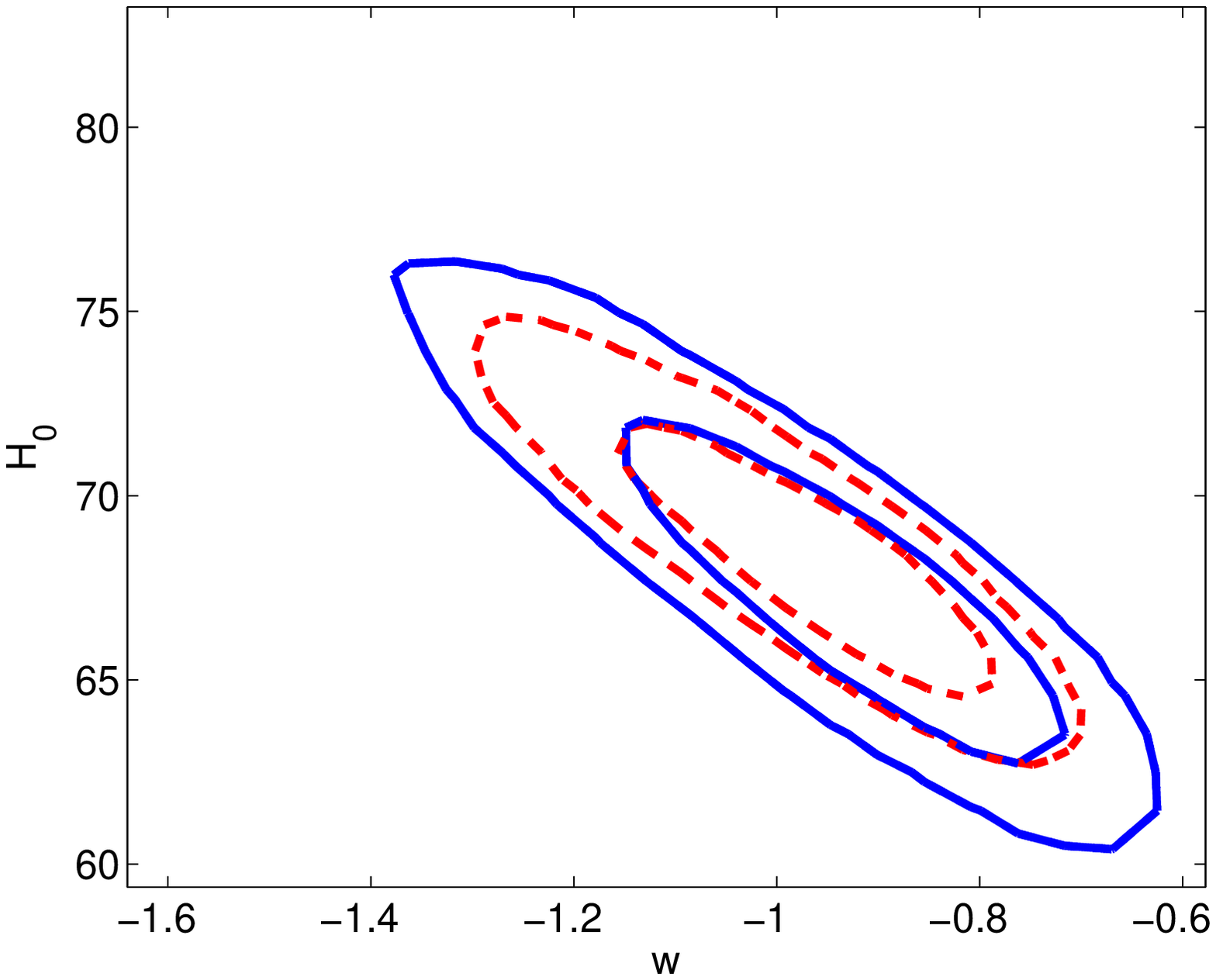}
\includegraphics[scale=0.31]{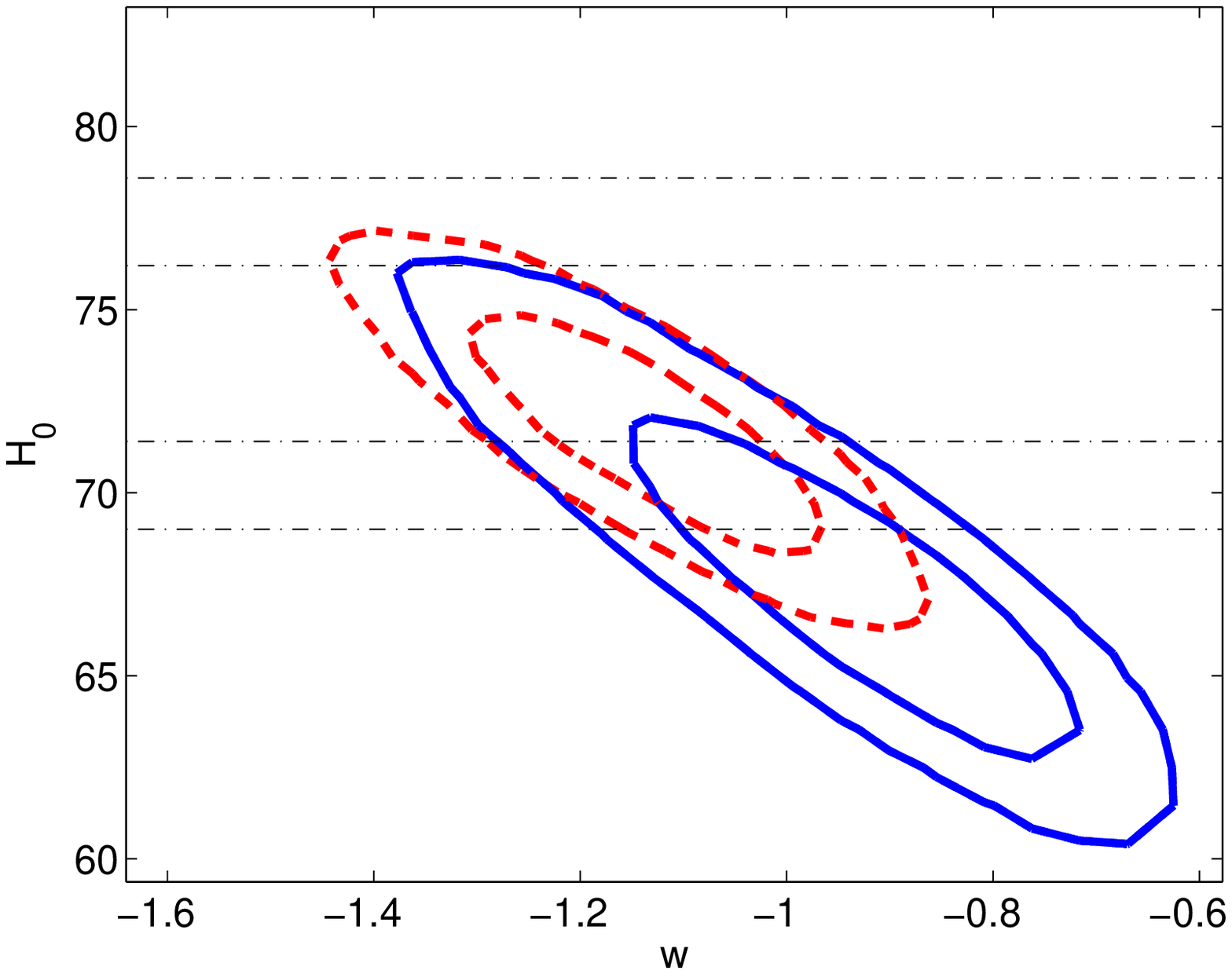}
\caption{Two-dimensional marginalized distribution between $w$ and $H_0$. The blue solid lines are obtained from WMAP9+BAO, and the red dotted lines denotes the constraints from WMAP9+BAO+$Hz~data$ (left panel) and WMAP9+BAO+HST $H_0$ (right panel), respectively. Four horizontal dashed lines denote the $1,2\,\sigma$ limits of the HST $H_0$ prior.}\label{fig_w_2d}
\end{center}
\end{figure}

Due to the degeneracy, WMAP9 data alone can not constrain $w$ very well. The efficient way to improve the limit is to add the SNIa observation at low redshifts. When using WMAP9 and ``Union2.1'' data, the constraint on $w$ is much more stringent, $w=-1.011^{+0.074}_{-0.073}$ ($68\%$ C.L.). Comparing with the SNIa data, the constraining power of the ``$Hz~data$'' is weaker. Adding the ``$Hz~data$'' can not improve the constraint too much, namely the $68\%$ C.L. constraint $w=-0.997\pm{0.072}$. In order to illustrate the constraining power of the $H(z)$ data better, in the following analysis we choose the WMAP9+BAO data combination as an example, instead of the WMAP9+SNIa data. WMAP9+BAO data combination yields the $68\%$ constraint on the constant EoS of dark energy of $w=-0.96\pm{0.15}$. When we include the ``$Hz~data$'' into the calculations, the limit on $w$ is slightly improved,
\begin{equation}
w=-0.98^{+0.12}_{-0.13}~~(68\%~{\rm C.L.})~.
\end{equation}
In the left panel of Figure \ref{fig_w_2d} we show the two-dimensional constraints in the ($w,H_0$) panel from WMAP9+BAO and WMAP9+BAO+$Hz~data$. Our results are similar to the limit from previous works (see e.g. refs. \cite{xia2008}).

The direct $H_0$ prior also affects the constraints of the constant dark energy EoS, since $w$ and $H_0$ are strongly anti-correlated which is clearly shown in Figure \ref{fig_w_2d}. Because the HST prior gives a slightly high value of $H_0$, WMAP9+BAO+HST $H_0$ combination yield the constraint on the constant EoS of $w=-1.14\pm{0.11}$ at $68\%$ confidence level, which mildly favors the phantom model with $w<-1$. One can see that the limits of the constant EoS are strongly dependent on our HST $H_0$ prior. However, the current observational data is still consistent with the standard $\Lambda$CDM model $w=-1$, due to the limits of the precisions of observational data.

\subsection{Parameters associated with Primordial Perturbations}

Inflation, the most attractive paradigm in the very early universe, has successfully resolved many problems existing in hot big bang cosmology, such as the flatness, horizon, monopole problem, and so forth \cite{inflation1,inflation2,inflation3}. Its quantum fluctuations turn out to be the primordial density fluctuations which seed the observed large scale structures and the anisotropies of CMB. Inflation theory has successfully passed several nontrivial tests. Currently, the cosmological observational data are in good agreement with a Gaussian, adiabatic, and scale-invariant primordial spectrum, which are consistent with single-field slow-roll inflation predictions. In this subsection, we discuss the constraints on the inflationary parameter, the tensor-to-scalar ratio $r$ and the running of spectral index $\alpha_s$, from the current data. As we know, the Hubble parameter is not related to the inflationary parameters straightforwardly. But the additional observations of Hubble parameter could improve constraints of some other parameters, like $\Omega_m$, which could affect the limits of inflationary parameters indirectly. Therefore, it is still very interesting to investigate the effect of the $H(z)$ data on the constraints. The numerical results on $r$ and $\alpha_s$ are shown in Table \ref{table_inflation}.

\begin{table*}[t] 
\caption{$68\%$ constraints on the tensor-to-scalar ratio $r$ and the running spectral index $\alpha_s$ and some other cosmological parameters from different data combinations. For the tensor-to-scalar ratio, we quote the $95\%$ upper limits instead.}\label{table_inflation}
\begin{center}
\begin{tabular}{c|c |c| c|c |c |c|c| c| c}
\hline
& \multicolumn{3}{c|}{WMAP9}& \multicolumn{3}{c|}{WMAP9+SNIa}& \multicolumn{3}{c}{WMAP9+BAO} \\
\cline{2-10}
& $r$ & $H_0$ & $\Omega_m$ & $r$ & $H_0$ & $\Omega_m$ &$r$ & $H_0$ & $\Omega_m$\\
\hline
+&$<0.46$&$73.7^{+3.2}_{-3.1}$&$0.245^{+0.029}_{-0.030}$&$<0.33$&$72.4^{+2.3}_{-2.2}$&$0.255^{+0.022}_{-0.023}$&$<0.19$&$68.91^{+0.98}_{-0.99}$&$0.293\pm{0.012}$ \\
\hline
+$Hz~data$&$<0.28$&$71.3\pm{1.7}$&$0.267\pm{0.019}$&$<0.26$&$71.2\pm{1.5}$&$0.268^{+0.016}_{-0.017}$&$<0.19$&$69.15^{+0.84}_{-0.86}$&$0.291\pm{0.010}$\\
\hline
+HST $H_0$&$<0.36$&$73.6^{+1.9}_{-1.8}$&$0.244\pm{0.018}$&$<0.33$&$73.0\pm{1.6}$&$0.249\pm{0.016}$&$<0.21$&$69.61^{+0.92}_{-0.93}$&$0.286^{+0.010}_{-0.011}$\\
\hline\hline
\hline
& \multicolumn{3}{c|}{WMAP9}& \multicolumn{3}{c|}{WMAP9+SNIa}& \multicolumn{3}{c}{WMAP9+BAO} \\
\cline{2-10}
& $100\alpha_s$ & $H_0$ & $\Omega_m$ & $100\alpha_s$ & $H_0$ & $\Omega_m$ &$100\alpha_s$ & $H_0$ & $\Omega_m$\\
\hline
+&$-1.4\pm{2.4}$&$69.1\pm{3.5}$&$0.295^{+0.045}_{-0.043}$&$-0.9\pm{2.1}$&$69.8\pm{2.4}$&$0.283\pm{0.028}$&$-1.9\pm{1.7}$&$68.3\pm{1.0}$&$0.301\pm{0.012}$ \\
\hline
+$Hz~data$&$-1.0\pm{1.9}$&$69.8\pm{1.8}$&$0.284\pm{0.022}$&$-0.8\pm{1.7}$&$70.0\pm{1.6}$&$0.281\pm{0.019}$&$-1.8\pm{1.7}$&$68.66^{+0.94}_{-0.95}$&$0.297\pm{0.012}$\\
\hline
+HST $H_0$&$0.5\pm{1.9}$&$72.2^{+2.0}_{-2.1}$&$0.256\pm{0.022}$&$0.1\pm{1.9}$&$71.9^{+1.7}_{-1.8}$&$0.260\pm{0.018}$&$-1.4^{+1.7}_{-1.6}$&$69.08\pm{0.98}$&$0.292\pm{0.012}$\\
\hline
\end{tabular}
\end{center} 
\end{table*}

Firstly, we consider the constraint on the tensor-to-scalar ratio $r$. In the standard $\Lambda$CDM model, the primordial tensor fluctuations could contribute to the CMB temperature and polarization power spectra. The most direct way of testing for a tensor contribution is to search for the large-scale B-mode pattern in CMB polarization which is very difficult to detect. Therefore, the amplitude of tensor modes is usually constrained from the measurements of the CMB temperature power spectrum. WMAP9 data alone yield the $95\%$ upper limit on the tensor-to-scalar ratio of $r<0.46$. When we add the SNIa data, the limit on $r$ is tighter, namely $r < 0.33$ ($95\%$ C.L.). The assumption that primordial fluctuations are purely scalar modes is still supported by the data.

\begin{figure}[t]
\begin{center}
\includegraphics[scale=0.28]{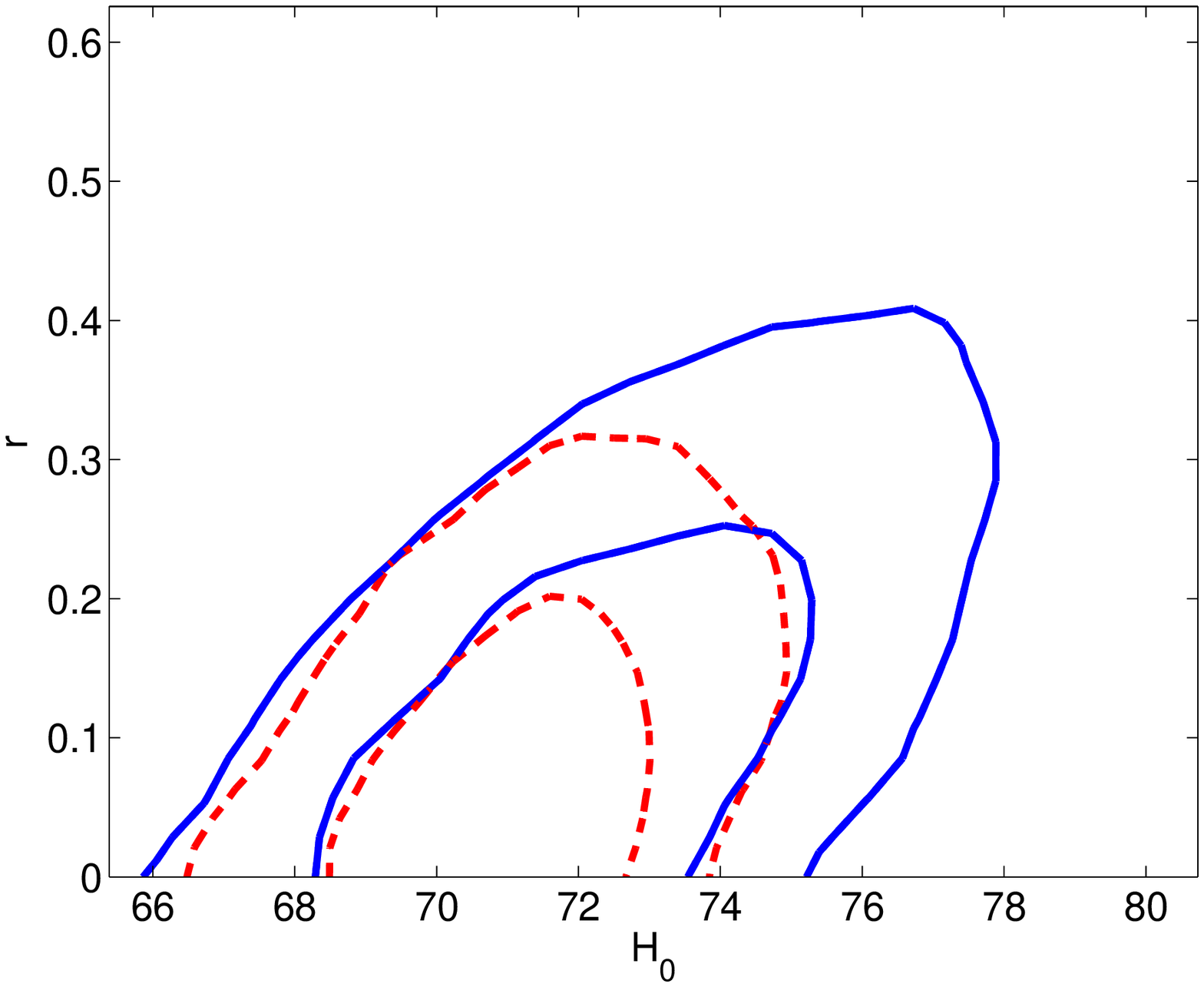}
\includegraphics[scale=0.27]{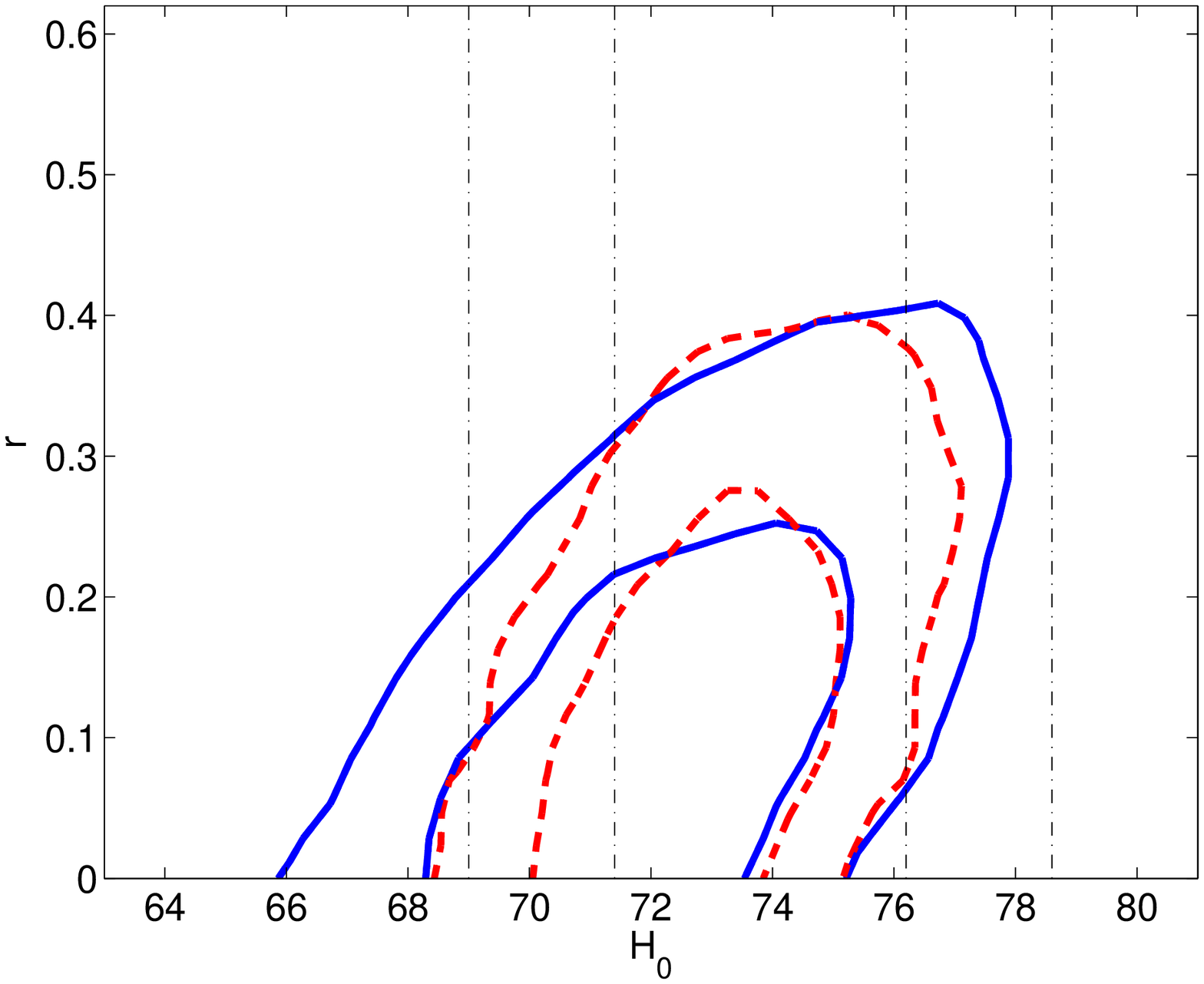}
\includegraphics[scale=0.31]{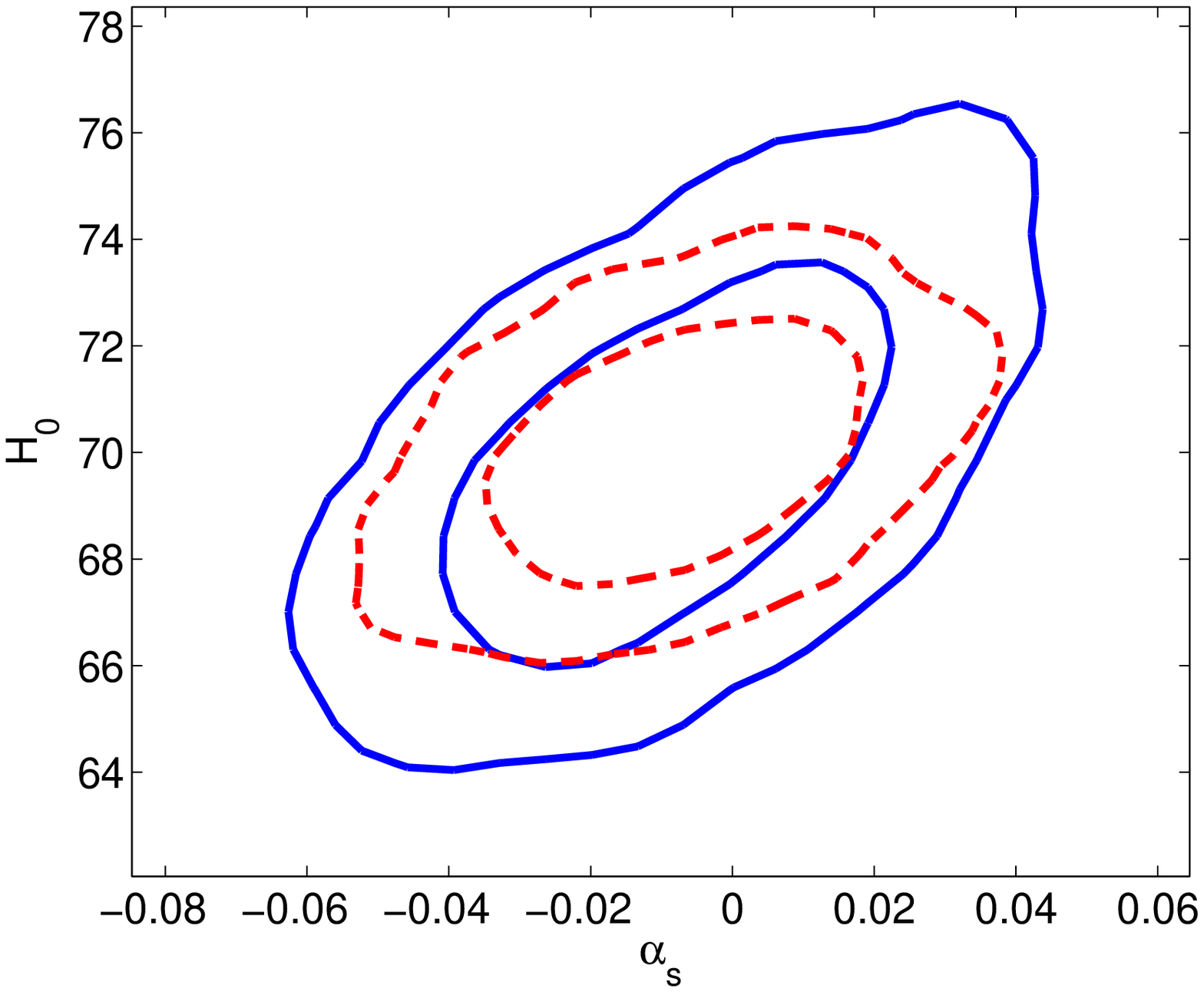}
\includegraphics[scale=0.31]{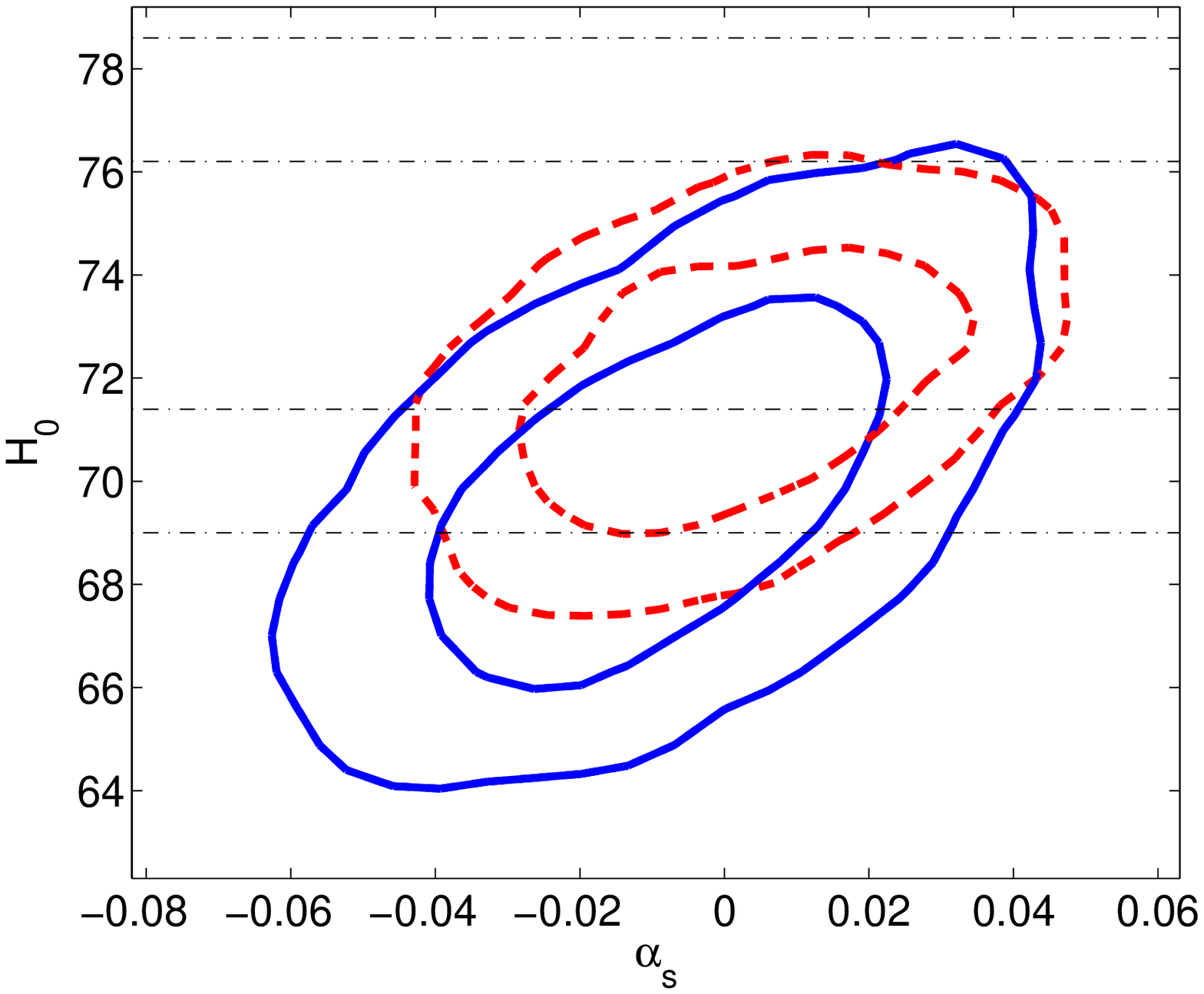}
\caption{Two-dimensional marginalized distributions in the ($H_0,r$) and ($H_0,\alpha_s$) panels. The blue solid lines are obtained from WMAP9+SNIa, and the red dotted lines denotes the constraints from WMAP9+SNIa+$Hz~data$ (left panels) and WMAP9+SNIa+HST $H_0$ (right panels), respectively. Four dashed lines denote the $1,2\,\sigma$ limits of the HST $H_0$ prior.}\label{fig_inf_2d}
\end{center}
\end{figure}

Since the Hubble parameter is related to the inflationary parameter indirectly, when including the ``$Hz~data$'', the constraint on $r$ is slightly improved,
\begin{equation}
r<0.26~~(95\%~{\rm C.L.})~.
\end{equation}
In the upper two panels of Figure \ref{fig_inf_2d}, we can see that adding the ``$Hz~data$'' improves the constraints on the current Hubble constant $H_0$ and reduces the correlations between $r$ and $H_0$. Therefore, we obtain tighter constraint on $r$ than WMAP9+SNIa data. We also investigate the effect of the HST $H_0$ prior on the constraint of $r$. We find the constraint on $r$ is almost identical with that from WMAP9+SNIa. This is because that the WMAP9+SNIa data yield the similar constraint on the Hubble constant with that of the HST prior, $H_0=72.4^{+2.3}_{-2.2}$ ${\rm km\,s^{-1}\,Mpc^{-1}}$ ($68\%$ C.L.). The HST prior does not provide extra information on $H_0$.

Finally, we explore the constraint on the running of the spectral index from the latest observational data. The simplest single-field inflationary models predict that the running of the spectral index should be of second order in inflationary slow-roll parameters and therefore small \cite{small_as}. Nevertheless, it is easy to construct inflationary models that have a larger scale dependence. So it is instructive to use the current data to constrain running of the spectral index $\alpha_s\equiv dn_s/d\ln{k}$.

Using the WMAP9 data alone, we do not find the significant running, $\alpha_s=-0.014\pm{0.024}$ at $68\%$ confidence level. Combining the WMAP9 data with the ``Union2.1'' data, the constraint on $\alpha_s$ is slightly tighter, $\alpha_s=-0.009\pm{0.021}$ ($68\%$ C.L.). When we add the ``$Hz~data$'' into the calculation, the data yield the $68\%$ C.L. constraint on the running of the spectral index of $\alpha_s=-0.008\pm{0.017}$. In the lower two panels of Figure \ref{fig_inf_2d}, we show the two-dimensional contours in the ($H_0,\alpha_s$) panels from different data combinations. There is a positive correlation between $\alpha_s$ and $H_0$. Therefore, when we use the HST prior, due to the effect of higher $H_0$ value, the obtained constraint on $\alpha_s$ is higher than that from without $H_0$ prior, namely the $68\%$ C.L. limit are $\alpha_s=0.001\pm{0.019}$ and $\alpha_s=-0.009\pm{0.021}$ from WMAP9+SNIa+HST $H_0$ and WMAP9+SNIa, respectively. The current data still gives no support for a running spectral index.

\section{Summary}\label{summary}

In this paper, we study the constraints on several cosmological parameters from the measurements of Hubble parameter, as well as the WMAP9, ``Union2.1'' compilation and some measurements of BAO. Here, we consider six extensions to the standard $\Lambda$CDM model, the constant EoS of dark energy $w$, the curvature of the universe $\Omega_k$, the total neutrino mass $\sum{m_\nu}$, the effective number of neutrinos $N_{\rm eff}$, and the parameters associated with the power spectrum of primordial fluctuations $\alpha_s$, $r$.

In order to investigate the constraining power of measurements of Hubble parameters data, we consider two kinds of $H(z)$ data sets: the ``$Hz~data$'' with 25 samples obtained from refs. \cite{gaztanaga,stern,moresco,simon,zhang2012} and the HST $H_0$ prior. Since the Hubble parameters data could provide the extra information of the expansion rate of universe at late times, when including the ``$Hz~data$'' into the analyses, it is very helpful to break some degeneracies among cosmological parameters. Therefore, we obtain better constraints on these parameters.

The HST measurement on the Hubble constant, HST $H_0$ prior, can obviously impact on the median values of parameters and also give good constraints on some of them, such as the effective number of neutrinos $N_{\rm eff}$, due to the degeneracies between these parameters and $H_0$.

Overall, our study shows that the observations on the Hubble parameter, which include direct measurements at different redshifts and today's Hubble constant, can potentially to be a complementary probes to other astronomy observations, such as the CMB, SNIa and large scale structure in modern cosmology.

\section*{Acknowledgements}

HL is supported in part by the National Science Foundation of China under Grant Nos. 11033005 and 11322325, by the 973 program under Grant No. 2010CB83300, by the Chinese Academy of Science under Grant No. KJCX2-EW-W01. JX is supported by the National Youth Thousand Talents Program and the grants No. Y25155E0U1 and No. Y3291740S3. ML is supported in part by National Science Foundation of China under Grants Nos. 11075074 and 11065004.


\end{document}